\documentclass[prl,aps,twocolumn,showpacs]{revtex4}
\usepackage{graphicx,latexsym}
\usepackage{dcolumn}
\usepackage{amsmath,amssymb,epsf,bm}
\begin{document}
\title{Spin cloud induced around an elastic scatterer by the intrinsic spin-Hall effect}
\author{A.~G. Mal'shukov$^1$, C.~S. Chu$^{2}$}
\affiliation{$^1$Institute of Spectroscopy, Russian Academy of
Science, 142190, Troitsk, Moscow oblast, Russia \\
$^2$Department of Electrophysics, National Chiao Tung University,
Hsinchu 30010, Taiwan}
\begin{abstract}
Similar to the Landauer electric dipole created around an impurity
by the electric current, a spin polarized cloud of electrons can be
induced by the intrinsic spin-Hall effect near a spin independent
elastic scatterer. It is shown that in the ballistic range around
the impurity, such a cloud appears in the case of Rashba spin-orbit
interaction, even though the bulk spin-Hall current is absent.
\end{abstract}
\pacs{72.25.Dc, 71.70.Ej, 73.40.Lq}

\maketitle

The spin-Hall effect attracts much interest because it provides a
method for manipulating electron spins by electric gates,
incorporating thus spin transport into conventional semiconductor
electronics. As it has been initially predicted, the electric field
$\mathbf{E}$ induces the spin flux of electrons or holes flowing in
the direction perpendicular to $\mathbf{E}$. This spin flux can be
due either to the intrinsic spin-orbit interaction (SOI) inherent to
a crystalline solid \cite{Murakami}, or to spin dependent scattering
from impurities \cite{Hirsch}. Intrinsic spin-Hall effect
corresponding to the former situation has been observed in p-doped
2D semiconductor quantum wells \cite{Wunderlich}, while the
extrinsic effect related to the latter scenario has been detected in
n-doped 3D semiconductor films \cite{Awschalom2}.

Most of the theoretical studies on the spin-Hall effect (SHE) has
been focused on calculation of the spin current (for a review see
\cite{Engel}). On the other hand, since the spin current carries the
spin polarization, one would expect a buildup of the spin density
near the sample boundaries. Such a spin accumulation near interfaces
of various nature was calculated in a number of works
\cite{Accumulation,Accumulation2,ballistic}. This accumulated
polarization is a first evidence of SHE that has been observed
experimentally in Ref. \cite{Wunderlich,Awschalom2}. In fact,
measuring spin polarization is thus far the only practical way to
detect SHE.

Yet the spin accumulation near interfaces is not the only signature
of SHE. To draw an analogy with the charge transport, one can expect
that similar to Landauer charge dipoles created by the DC electric
current around impurities \cite{Landauer}, nonequilibrium spin
dipoles must be formed subsequent to the spin-Hall current. One may
expect that the spin cloud will appear around a spin-orbit scatterer
in case of extrinsic SHE, as well as around a spin-independent
scatterer, in case of the intrinsic effect. We will consider the
latter possibility for a 2D electron gas with Rashba interaction.
The polarization in the direction perpendicular to 2DEG will be
calculated in the ballistic range around an impurity represented by
an isotropic spin independent scattering potential. Besides
conventional semiconductor quantum wells this analysis can be
applied to metal adsorbate systems with strong Rashba type spin
splitting of surface states \cite{metalRashba}. In this case the
spin cloud can be measured by STM with a magnetic tip.

The Landauer electric dipole has been calculated \cite{Chu,Zwerger}
basing on the asymptotic expansion of the electron waves elastically
scattered from an isolated impurity. Subsequent averaging of the
corresponding spatial probability weighted by the Boltzmann
distribution function of incident wavevectors produces the dipole
distribution. The spin cloud could be obtained in a similar way.
Instead, we choose a Green function method combined with the linear
response theory. Within this method the spin density is given by the
standard Kubo formula where the scattering potential of a target
impurity, at a fixed position $\mathbf{r}_i$, is included into the
Green functions, up to the second perturbation order. Other
impurities are assumed to be randomly distributed over a 2D plane,
so that the calculated spin density is averaged over their
positions.

We assume that a uniform external electric field is applied to 2DEG.
The field is represented by the vector potential $\mathbf{A}$,
$\mathbf{E}=i\omega\mathbf{A}/c$, with $\omega \rightarrow$ 0 in the
DC regime. The corresponding interaction Hamiltonian is
$e\mathbf{A\cdot v}/c$, where the velocity $\mathrm{v}^j$, $j=x,y$,
includes the spin-orbit correction $\partial
(\mathbf{h}_{\mathbf{k}}\cdot \bm{\sigma})/\partial k^j$. The
spin-orbit field $\mathbf{h}_{\mathbf{k}}$ is a function of the
two-dimensional wave-vector $\mathbf{k}$. In its turn, the
spin-orbit interaction is written in the form
\begin{equation}\label{Hso}
H_{so} = \mathbf{h}_{\mathbf{k}}\cdot\bm{\sigma} \, ,
\end{equation}
where $\bm{\sigma}$$\equiv$$(\sigma^x,\sigma^y,\sigma^z)$ is the
Pauli matrix vector. We assume that the target impurity, located at
$\mathbf{r}_i$, has a scattering potential
$U(\mathbf{r}-\mathbf{r}_i)$. In 2D geometry the corresponding Born
amplitude is given by
\begin{equation}\label{f}
f(\mathbf{k},\mathbf{k'})=-\frac{m^*}{\sqrt{2\pi k_F}}\int dr^2
U(\mathbf{r})e^{i(\mathbf{k}-\mathbf{k'})\mathbf{r}}\, ,
\end{equation}
where $\hbar=1$ and $\varphi$ is the angle between $\mathbf{k}$ and
$\mathbf{k'}$. Both the scattered and the incident wavevectors are
taken at the Fermi circle with the radius $k_F$. Other impurities,
which not necessarily are of the same nature as the target impurity,
are randomly distributed within a sample. They create the random
potential $V_{sc}(\mathbf{r})$ which is assumed to be delta
correlated, so that the pair correlator $\langle
V_{sc}(\mathbf{r})V_{sc}(\mathbf{r}^{\prime})\rangle=\Gamma
\delta(\mathbf{r}-\mathbf{r}^{\prime})/\pi N_F$, where $N_F$ is the
electron density of states at the Fermi energy, and $\Gamma=1/2\tau$
is expressed via the mean elastic scattering time $\tau$. Assuming
that the semiclassical approximation $E_F\tau \gg 1$ is valid, one
can apply the standard perturbation theory \cite{agd,alt} when
calculating the configurational averages of Green functions and
their products. In the leading order of $(E_F\tau)^{-1}$ and up to
the second order in $U(\mathbf{r}-\mathbf{r}_i)$, the average
retarded Green function in the momentum representation is given by
\begin{equation}\label{G}
G^r_{\mathbf{k}\mathbf{k'}}(\omega)=\delta_{\mathbf{k}\mathbf{k'}}G^{r(0)}_{\mathbf{k}}(\omega)+
G^{r(1)}_{\mathbf{k}\mathbf{k'}}(\omega)+G^{r(2)}_{\mathbf{k}\mathbf{k'}}(\omega)\,
,
\end{equation}
with the unperturbed function given by the 2$\times$2 matrix
\begin{equation}\label{G0}
 G^{r(0)}_{\mathbf{k}}(\omega)
 = (\omega - E_{\mathbf{k}} -
\bm{h}_{\mathbf{k}}\cdot\bm{\sigma} + i\Gamma)^{-1} \, ,
\end{equation}
where $E_{\mathbf{k}}$=$k^2/(2m^*)$. Other functions in (\ref{G})
are
\begin{eqnarray}\label{G12}
 G^{r(1)}_{\mathbf{k}\mathbf{k'}}(\omega)&=&
 G^{r(0)}_{\mathbf{k}}(\omega)U_{\mathbf{k}\mathbf{k'}}G^{r(0)}_{\mathbf{k'}}(\omega)
  \\
G^{r(2)}_{\mathbf{k}\mathbf{k'}}(\omega)&=&
 G^{r(0)}_{\mathbf{k}}(\omega)\sum_{\mathbf{k''}}
 U_{\mathbf{k}\mathbf{k''}}G^{r(0)}_{\mathbf{k''}}(\omega)
 U_{\mathbf{k''}\mathbf{k'}}G^{r(0)}_{\mathbf{k'}}(\omega)\, .
 \nonumber
\end{eqnarray}
The matrix elements $U_{\mathbf{k'}\mathbf{k}}= -\sqrt{2\pi
k_F}f(\mathbf{k},
\mathbf{k'})\exp[i(\mathbf{k}-\mathbf{k'})\mathbf{r}_i]/m^*$.
Expressions similar to Eqs.(\ref{G}-\ref{G12}) can be obtained for
the advanced functions
$G^a_{\mathbf{k'}\mathbf{k}}(\omega)=G^r_{\mathbf{k'}\mathbf{k}}(\omega)^{\dag}$.
The sum over $\mathbf{k''}$ in the second Eq.~(\ref{G12}) can be
directly calculated. First, we decompose $G^{r(0)}_{\mathbf{k''}}$
into a spin independent scalar part and a spin dependent part which
is proportional to $\mathbf{h}_{\mathbf{k''}}\cdot\bm{\sigma}$. Due
to the time inversion symmetry
$\mathbf{h}_{\mathbf{k''}}=-\mathbf{h}_{-\mathbf{k''}}$ the sum over
$\mathbf{k''}$ on the spin dependent part is zero for an isotropic
scattering amplitude. For anisotropic amplitude, however, this sum
is not identically 0. Nevertheless, the sum on the spin dependent
part can be ignored either way in the following calculations,
because it is proportional to the small parameter $h_{k_F}/E_F \ll
1$. Further, it is easily seen that only
$\mathrm{Im}[G^{r(0)}_{\mathbf{k''}}]$ is important in this
$\mathbf{k''}$ sum because the real part gives rise to a term that
simply adds to $U_{\mathbf{k}\mathbf{k'}}$ in the first line of
Eq.~(\ref{G12}), thus effectively renormalizing the Born scattering
amplitude. The imaginary part can not be absorbed in such a way
because it has opposite signs for the advanced and retarded Green
functions. Taking into account that $\omega \simeq E_F$ and assuming
that $h_{k_F} \ll E_F$ we thus get
\begin{eqnarray}\label{S}
&&\sum_{\mathbf{k''}}
U_{\mathbf{k}\mathbf{k''}}G^{r(0)}_{\mathbf{k''}}(\omega)
U_{\mathbf{k''}\mathbf{k'}}=-i \pi N_{F}S(\mathbf{k},\mathbf{k'})e^{i(\mathbf{k'}-\mathbf{k})\mathbf{r}_i}\, \nonumber \\
&&S(\mathbf{k},\mathbf{k'})=\frac{k_F}{m^{*2}}\int d\varphi^{\prime
\prime} f(\mathbf{k''},\mathbf{k}) f(\mathbf{k'},\mathbf{k''})\,,
\end{eqnarray}
where $\phi''$ is the angle of the vector $\mathbf{k''}$, with
$|\mathbf{k''}|=k_F$. At $\mathbf{k}=\mathbf{k'}$ the integral in
(\ref{S}) is equal to the scattering cross-section.

Within the semiclassical theory we follow the well known method
\cite{agd,alt} to calculate the configurational average of the
Green functions product that enters into the Kubo's linear
response equation. Because of scattering on a target impurity this
product contains more than a pair of Green functions. As our
leading approximation we take into account the so called ladder
perturbation series describing particle and spin diffusion
processes. When averaging the Green function product, within this
approximation, only pairs of retarded and advanced functions
carrying close enough momenta should be chosen to become elements
of the ladder series. This considerably simplifies calculations.
Some of the representative diagrams are shown at Fig.~1, where
$\mathbf{v}$ denotes the velocity operator
\begin{equation}\label{v}
\mathrm{v}^j=\frac{k^j}{m^*}+\frac{\partial
\mathbf{h}_{\mathbf{k}}\cdot \bm{\sigma}}{\partial k^j}\,.
\end{equation}
The diffusion ladder renormalizes only the vertex associated with
the electric field, while such a ladder, as we just explained,
does not appear at the vertex attributed to the induced spin
density. It is because in the ballistic range around the impurity,
the momentum transfer $|\mathbf{p}-\mathbf{k}| \gg 1/(v_F \tau)$,
and thus the diffusion is not important. Finally, the density of
spins oriented in $z$ direction can be written as
\begin{figure}[tp]
\includegraphics[width=7.5cm, height=4.5cm]{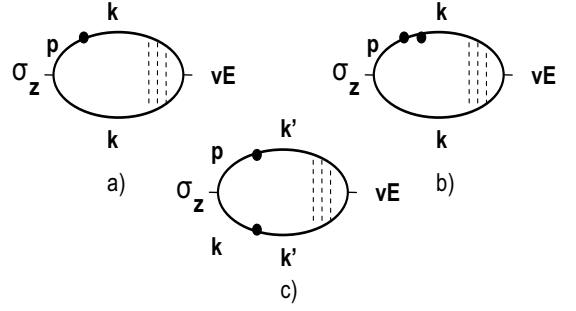}
\caption{Diagram for the spin density. Scattering of electrons by a
target impurity is shown by the solid circles. Dashed lines denote
the ladder series of particle scattering by the random potential.
$\mathbf{p}, \mathbf{k}, \mathbf{k'}$ are electron momenta.}
\label{fig1}
\end{figure}
\begin{eqnarray}\label{sigma}
\sigma_z(\mathbf{r})&=&\sum_{\mathbf{k},\mathbf{k}^{\prime},\mathbf{p}}
e^{i(\mathbf{p}-\mathbf{k})\mathbf{r}}\int
\frac{d\omega}{2\pi} \frac{d n_F(\omega)}{d\omega} \times \nonumber
\\
&& Tr[G^a_{\mathbf{k}^{\prime}\mathbf{k}}(\omega)\sigma_z
G^r_{\mathbf{p}\mathbf{k}^{\prime}}(\omega) \mathcal{T}(\omega
,\mathbf{k'})] \,,
\end{eqnarray}
where the trace runs through the spin variables and $n_F(\omega)$
is the Fermi distribution function. The functions
$G^{r/a}_{\mathbf{k}^{\prime}\mathbf{k}}$ are given by
Eq.~(\ref{G}). In (\ref{sigma}), only terms up to the second order
in $U_{\mathbf{k'}\mathbf{k}}$ should be taken into account. Hence
the highest order corrections are those shown at Fig.1, b) and c).
At low temperatures only $\omega$ in close vicinity around $E_F$
contributes to the integral in (\ref{sigma}). Therefore, below we
set $\omega=E_F$.

It is important that the vertex $\mathcal{T}(\omega ,\mathbf{k})$
in Eq.(\ref{sigma}) is the same that enters into the spin-Hall
conductance. On the other hand, as was shown in many publications
\cite{Inoue}, for linear in $\mathbf{k}$ SOI and $h_{\mathbf{k}}
\ll E_F$, a contribution to $\mathcal{T}(\omega ,\mathbf{k})$ from
the spin dependent part of the velocity (\ref{v}) cancels with its
spin independent part, while such a cancellation does not take
place in case of nonlinear SOI \cite{MalshDress}. As a result of
such cancellation, in a linear case $\mathcal{T}(\omega
,\mathbf{k})$ is reduced to the simple expression
\begin{equation}\label{TRashba}
\mathcal{T}(\omega ,\mathbf{k})=\frac{e}{m^*}\mathbf{k}\mathbf{E}
\,.
\end{equation}

Let us consider the spin density (\ref{sigma}) in the presence of
the Rashba spin-orbit field $h_x=\alpha k_y, h_y=-\alpha k_x$. In
the zeroth order in $U_{\mathbf{k}\mathbf{k'}}$ the Green
functions in (\ref{sigma}) are given by the first term in
(\ref{G}). In this approximation and with $\mathcal{T}$ given by
(\ref{TRashba}) one can easily see that $\sigma_z(\mathbf{r})=0$.
On the other hand, the inplane spin polarization directed
perpendicular to $\mathbf{E}$ is finite. This polarization is due
to the electric orientation effect \cite{Edelstein}. In the first
order with respect to $U_{\mathbf{k}\mathbf{k'}}$ the z spin
polarization is represented by Fig.1a. Expressing
$U_{\mathbf{k}\mathbf{k'}}$ via the scattering amplitude, from
(\ref{G}-\ref{G12}) and (\ref{sigma}) we obtain
\begin{eqnarray}\label{sigmaa}
\sigma_z^{(a)}(\mathbf{r})&=&-e\sqrt{\frac{k_F}{2\pi}}\sum_{\mathbf{k},\mathbf{p}}\frac{\mathbf{k}\mathbf{E}}{m^{*2}}
Tr[G^{r0}_{\mathbf{k}}(E_F)G^{a0}_{\mathbf{k}}(E_F) \times
\nonumber \\
&&\left(\sigma_z
G^{r0}_{\mathbf{p}}(E_F)f(\mathbf{p},\mathbf{k})e^{i(\mathbf{p}-\mathbf{k})\mathbf{R}}+h.c.\right)
] \,,
\end{eqnarray}
where $\mathbf{R}=\mathbf{r}-\mathbf{r}_i$. At $kR,pR \gg 1$ the
2D angular integration in (\ref{sigmaa}) can be performed by
expansions around the saddle-points $(\mathbf{p}\mathbf{R}/pR)=\pm
1$ and $(\mathbf{k}\mathbf{R}/kR)=\pm 1$, that result in the
asymptotic expansion of $\sigma_z(\mathbf{r})$ at a large distance
from the impurity. The remaining integrals over $p$ and $k$ are
dominated by contributions from the poles of Green functions
(\ref{G0}). These poles are located at $k,p=\pm k_F\pm
L_{so}^{-1}\pm il^{-1}$, where $L_{so}=\hbar/m^*\alpha \gg
k_F^{-1}$ is the characteristic spin-orbit length and $l$ is the
mean free path. In the ballistic range $R \lesssim l$ one may
substitute the imaginary part $il^{-1}$ of the poles by $i\delta$
with $\delta\rightarrow 0$. Depending on combination of ~$\pm$~
signs of the poles, the scattering amplitude entering into
(\ref{sigmaa}) will coincide either with the forward scattering
amplitude $f(0)=f(k_F\mathbf{\hat{R}},k_F\mathbf{\hat{R}})$, or
with the backscattering amplitude
$f(\pi)=f(k_F\mathbf{\hat{R}},-k_F\mathbf{\hat{R}})$, where
$\mathbf{\hat{R}}=\mathbf{R}/R$ is the unit vector directed to the
observation point. Finally, we obtain from (\ref{sigmaa}) and
(\ref{G0})
\begin{eqnarray}\label{sigmaafin}
\sigma_z^{(a)}(\mathbf{r})&=&\frac{m^*}{R}\sqrt{\frac{2}{\pi^3
k_F}}v^j_d\left(\frac{\partial \mathbf{n}_R}{\partial
R^j}\times\mathbf{n}_R \right) \times \nonumber \\
&&\mathrm{Re}[f(\pi)e^{2ik_FR}]\sin^2\left(\frac{R}{L_{so}}\right)
\,,
\end{eqnarray}
where $\mathbf{v}_d=e\tau\mathbf{E}/m^*$ is the drift velocity.
The unit vector
$\mathbf{n}_R=\mathbf{h}_{k_F\mathbf{\hat{R}}}/|h_{k_F\mathbf{\hat{R}}}|$.
For Rashba interaction it is $n_R^x=\hat{R}^y,\,
n_R^y=-\hat{R}^x$.

In a similar way and with the use of Eq.(\ref{S}) one can
calculate the second order contribution to the spin density, that
is represented in Fig.~1b) and c). Assuming the electric field to
be applied in the $x$-direction, we get the final result, which is
as a sum of all diagrams in Fig.~1a)-c),
\begin{eqnarray}\label{sigmafin}
\sigma_z(\mathbf{r})&=&-\frac{m^*v_d\sigma_{t}}{2\pi^2 RL_{so}}\sin
\left(\frac{2R}{L_{so}}\right)\sin\theta +\nonumber
\\
&&\frac{m^*v_d}{2\pi^2 R^2}
\sin^2\left(\frac{R}{L_{so}}\right)\sin^3\theta\times \nonumber
\\
&&\left(\sigma_{tot}+\sqrt{\frac{8\pi}{k_F}}\mathrm{Re}[f(\pi)e^{2ik_FR}]\right)
 \,,
\end{eqnarray}
where $\sigma_{tot}$ and $\sigma_{t}$ are the total and transport
scattering cross sections, respectively, and $\theta$ is the angle
between the vector $\mathbf{R}$ and the x-axis. In order to check
our method we applied it to the calculation of the charge dipole,
whence $\sigma_z$ is substituted by 1 in Eq.~(\ref{sigma}). Ignoring
SOI we obtained the same result as in Ref.\cite{Zwerger}.

The explicit shape of the spin cloud is clearly seen from
Eq.~(\ref{sigmafin}). It consists of a dipole, oriented
perpendicular to the electric field, and a tripole. Similar to the
Landauer charge dipole distribution \cite{Zwerger}, the spin density
contains both slowly varying and fast Friedel oscillation
components. Important distinctions, however, are found in the
asymptotic behaviors. First, unlike the charge density, whose slow
asymptotic term is represented by monotonous $R^{-1}$ dependence,
the spin density oscillates with a period determined by the
spin-orbit precession length $\pi L_{so}$. Second, at smaller
distances $R\lesssim L_{so}$, the polarization decreases as
$R^{-2}$. It should be noted that this asymptotic form can not be
obtained by the method based on the conventional leading order
asymptotic expansion of the wave function, as it has been done in
\cite{Zwerger} for the Landauer dipole. It is because in 2D geometry
the corresponding scattered amplitude decreases as $1/\sqrt{R}$.
Accordingly, the probability density, which can be either the charge
or the spin density, will be proportional to $1/R$, not $1/R^2$.

When talking about asymptotic expression (\ref{sigmafin}), one
should not forget that it is valid only within the ballistic range
$R \lesssim l$. At larger distances the ballistic part of the spin
density decays as $exp(-R/l)$. On the other hand, outside the
ballistic range the spin diffusion becomes important. Spin
diffuses during the D'yakonov-Perel' \cite{dp} spin relaxation
time, up to the distance $\sim L_{so}$. Hence, the spin diffusion
must be taken into account at $R \gg l$, providing that the
spin-orbit coupling is not too strong, so that $L_{so} \gg l$. In
order to calculate the spin density in the diffusive range, the
ladder diagrams renormalizing the left hand vertex in Fig.~1
should be taken into account. An evident result to be expected in
this case is that the diffusion spin cloud with the size $\gg l$
will appear in addition to Eq.(\ref{sigmafin}). Due to the spin
relaxation, however, the spin density will decay exponentially at
$R\gg L_{so}$. This behavior is in sharp contrast to the power law
decreasing of the charge density \cite{Chu}. In the latter case,
the long-range $R^{-1}$ charge-density tails of many impurities
result in the macroscopic electric field which can be related to
the electric potential difference at the sample boundaries. This
was the main idea by Landauer - to associate impurities with
resistors which give rise to an overall potential drop at a given
current. Similarly, one would try to formulate the spin-Hall
effect in terms of the \emph{spin-Hall} resistance and spin
dependent chemical potential $\chi(\mathbf{r})$, defined as
$N_F\chi(\mathbf{r})=\sum_i \sigma_z(\mathbf{r}-\mathbf{r}_i)$. In
such a way an influence of disorder on SHE can be considered from
the microscopic point of view starting from individual impurities.
For example, similar approach has been employed in a semiclassical
analysis of the side jump contribution to the anomalous Hall
effect \cite{Sinitsyn}. Returning to the spin potential
$\chi(\mathbf{r})$, one can notice that due to exponential decay
in space of the spin cloud, the well converging sum over
impurities will produce, on average, a vanishing spin-Hall
chemical potential everywhere, except for the $R\sim L_{so}$ range
near the sample boundary. No such spin accumulation, on the other
hand, has been found near hard wall flanks of a 2D diffusive strip
of 2DEG with Rashba SOI \cite{Accumulation2}. At the same time the
finite accumulation was calculated in \cite{Accumulation} for
other boundary conditions. Probably, this means that the spin
density outside the ballistic range around an elastic scatterer is
finite, but the combined spin density produced by many impurities
will depend on the boundary conditions for spin diffusion.

In conclusion, for a 2DEG with Rashba spin-orbit interaction we
calculated the nonequilibrium spin polarization induced by the
intrinsic spin-Hall effect in the ballistic range around a spin
independent scatterer. The angular spatial distribution of the
spin density is represented by a tripole and a dipole oriented
perpendicular to the electric field. As a function of the distance
from the scatterer, the polarization shows the power law decay
with oscillations, some terms oscillating relatively slowly, with
the period $\pi L_{so}$, while other terms varying fast, with a
period of Friedel oscillations. Noteworthy, that although the
z-polarized spin-Hall current is zero in case of Rashba SOI, we
found out that the z-component of the spin density is not zero in
the ballistic range. This agrees with finite spin accumulation
near flanks of a ballistic strip \cite{ballistic}.

This work was supported by the Taiwan National Science Council
NSC94-2811-M-009-010 and RFBR Grant No 060216699. A.G.M.
acknowledges the hospitality of Taiwan National Center for
Theoretical Sciences.

\end{document}